\documentclass[useAMS,usenatbib,letter]{mn2e}
\usepackage{graphicx}

\def\etal{et~al. }
\def\msun{$M_\odot$}

\begin{document}

\title[Geometrical and Physical Properties of Circumbinary Discs in
Eccentric Stellar Binaries] {Geometrical and Physical Properties of
Circumbinary Discs in Eccentric Stellar Binaries}

 \author[B. Pichardo, L.S. Sparke \& L.A. Aguilar]{Barbara Pichardo$^{1,2}
$\thanks{E-mail:barbara@astroscu.unam.mx (BP);sparke@astro.wisc.edu(LSS);
aguilar@astrosen.unam.mx (LAA)}, Linda S. Sparke$^{3}$\footnotemark[1],
Luis A. Aguilar$^{4}$\footnotemark[1]\\
$^{1}$Instituto de Astronom\'\i a, Universidad Nacional Aut\'onoma de M\'
exico, Apdo. postal 70-264 Ciudad Universitaria, D.F., M\'exico\\
$^{2}$Institute for Theoretical Physics, University of Zurich, Winterthur
erstrasse 190, 8057, Switzerland\\
$^{3}$Department of Astronomy, 475 North Charter Street, University of
Wisconsin-Madison, Madison WI 53706-1582, USA\\
$^{4}$Observatorio Astron\'omico Nacional, Universidad Nacional Aut\'onom
a de M\'exico, Apdo. postal 877, 22800 Ensenada, M\'exico
}

\date{Accepted . Received ; in original form }

\pagerange{\pageref{firstpage}--\pageref{lastpage}} \pubyear{}

\maketitle

\label{firstpage}

\begin{abstract}
In a previous work (Pichardo \etal 2005), we studied stable
configurations for circumstellar discs in eccentric binary systems.
We searched for ``invariant loops'': closed curves (analogous to
stable periodic orbits in time-independent potentials) that change
shape with the binary orbital phase, as test particles in them move
under the influence of the binary potential. This approach allows us
to identify stable configurations when pressure forces are
unimportant, and dissipation acts only to prevent gas clouds from
colliding with one another.  We now extend this work to study the main
geometrical properties of circumbinary discs. We have studied more
than 100 cases with a range in eccentricity $0 \le e \le 0.9$, and
mass ratio $0.1 \le q \le 0.9$. Although gas dynamics may impose
further restrictions, our study sets lower stable bounds for the size
of the central hole in a simple and computationally cheap way, with a
relation that depends on the eccentricity and mass ratio of the
central binary. We extend our previous studies and focus on an
important component of these systems: circumbinary discs. The radii
for stable orbits that can host gas in circumbinary discs are sharply
constrained as a function of the binary's eccentricity.  The
circumbinary disc configurations are almost circular, with
eccentricity $e_d < 0.15$, but if the mass ratio is unequal the disk
is offset from the center of mass of the system.  We compare our
results with other models, and with observations of specific systems
like GG Tauri A, UY Aurigae, HD 98800 B, and Fomalhaut, restricting
the plausible parameters for the binary.
\end{abstract}


\section{Introduction}
\label{Intro}

It is currently believed that fragmentation is the most probable
mechanism for star formation, and the main product of fragmentation
are multiple stellar systems with preference for wide eccentric
binaries with separations $\ge 10 AU$ (Bonnell \& Bastien 1992; Bate
1997; Bate \& Bonnell 1997; Bodenheimer \etal 2000). Even in isolated
stars, there is evidence that the majority of Sun-like stars formed in
clusters (Carpenter 2000; Lada \& Lada 2003), including the Sun
(Looney, Tobin \& Fields 2006).

In the last decade the interest in binary systems has increased. This
is in part because of the discovery that many T-Tauri and other
pre-main sequence binary stars possess circumstellar and circumbinary
discs as inferred from observations of excess radiation at infrared to
millimeter wavelengths, polarization, and both Balmer and forbidden
emission lines (Mathieu \etal 2000, Itoh \etal 2002, for a review see
Mathieu 1994). On the other hand, recent observations of binary star
systems, using the Spitzer Space Telescope, show evidence of debris
discs in these environments (Trilling \etal 2007) and planets (Fischer
\etal 2008). In their studies they find that $60\%$ of the observed
close binary systems (separations smaller than 3 $AU$) have excess in
their thermal emission, implying on-going collisions in their
planetesimal regions.

Over 150 extrasolar planets have been identified in surveys using the
Doppler technique. Of the first 131 extrasolar planetary systems that
have been confirmed, at least 40 are in binary or multiple systems
(for an up-to-date list see Haghighipour 2006). Approximately 30 of
them are on S-type orbits (around one of the components: circumstellar
discs) with wide stellar separations (between 250 and 6500 $AU$),
including at least 3 that orbit one member of a triple star (Raghavan
\etal 2006). Although most of these binaries are very wide, a few have
separations smaller than 20 $AU$ (Els \etal 2001; Hatzes \etal 2003),
challenging standard ideas of Jovian planet formation. Some
interesting ideas try to explain the formation of Jovian planets
within close binaries, but they could only explain few cases (Pfahl \&
Muterspaugh 2006), if more are discovered soon, these theories would
not be sufficient. Although close binaries are not included in precise
Doppler radial velocity search programs because of their complex and
varying spectra, at least one planet with a minimum of $~2.5$ Jupiter
masses has been detected in a P-type orbit (around both components:
circumbinary discs), with a distance from the center of mass of $~23$
$AU$. The source is a radio pulsar binary comprised by a neutron star
and a white dwarf in a $~191$ day stellar orbit (Lyne \etal 1988;
Sigurdsson and Phinney, 1993; Sigurdsson \etal 2003). An example of
accretion in P-type orbits about close binaries is given by Quintana
\& Lissauer (2006) who note the observation of the two small moons
orbiting in nearly circular/planar orbits about the binary system
Pluto-Charon (Weaver \etal 2006). Specifically regarding to
circumbinary disc material, millimeter and mid-infrared excess
emission has been detected around several spectroscopic pre-main
sequence binary star systems including GW Ori (Mathieu \etal 1995), UZ
Tau E (Jensen \etal 1996), DQ Tau (Mathieu \etal 1997).

In this work we have followed the same steps as in Pichardo, Sparke \&
Aguilar (2005, hereafter Paper~I), where we opted for a simpler
approach, analogous to using the structure of periodic orbits in a
circular binary, to predict the gas flow. The path followed by a gas
parcel in a stable disk around a star must not intersect itself, or
the path of a neighboring parcel (unlike the case of planets, where
the paths may cross). In our work we follow Rudak \& Paczynski (1981)
and explore those non-crossing orbits of test particles that could be
interpreted as gas particles in the low pressure regime, or as
protoplanets or planets. An important issue in Celestial Mechanics is
to determine the regions around a stellar binary system where
accretion discs can form. Important theoretical effort carried out to
answer this question is reviewed in Paper~I, where we studied
circumstellar and circumbinary discs in binaries of arbitrary
eccentricity and mass ratio.  In this work we extend those studies,
which were based on identifying families of stable {\it invariant
loops}, a concept introduced by Maciejewski \& Sparke (1997, 2000) in
studies of nested galactic bars. We focus this time specifically on
the geometry of circumbinary discs. We employ for this approach a test
particle method probing the orbital structure of binaries of various
eccentricities and mass ratios.

In Section \ref{method} we briefly review the concept of an {\it
invariant loop}, describe the method to solve the motion equations and
the strategy used to find invariant loops. The geometry of the
circumbinary discs including a fit for the inner radii of the
circumbinary disc and a fit for the lopsidedness, are presented in
Section \ref{geom}. In Sections \ref{theory} and \ref{observations} we
apply this study to compare with theoretical work and observations of
some well known systems, respectively. Our conclusions are presented
in section \ref{conclusions}.

\section{The Method and Numerical Implementation}\label{method}

A more detailed description of the invariant loops method and its
numerical implementation is given in Paper~I (also in Maciejewski \&
Sparke -1997, 2000-). We give in this section a brief description.

In the well studied circular 3-body problem, one known integral of
motion is conserved: the Jacobi constant, defined in the rotating
reference frame of the stars. Stable periodic orbits in this rotating
system are defined and represent the ``backbone'' of the orbital
structure. On the other hand, when the eccentricity is non-zero, there
are no known integrals of motion to facilitate any analytical studies.
However, the lack of a global integral of motion, does not preclude
the existence of restrictions that apply to particular orbits. For
motion in the plane of the binary, an additional integral of motion
would confine an orbit to lie on a 1-dimensional curve every time the
system comes back to the initial orbital phase.  For example, if we
look at the system every time the binary is at periastron, a particle
following this orbit will land in a different spot but on the same
1-dimensional curve, which we call an {\it invariant loop}.  In this
manner {\it invariant loops} represent the generalization of periodic
orbits for periodically time-varying potentials. This means that an
invariant loop is not a simple orbit but an ensemble of orbits that
lie on a 3-torus in this extended phase-space, but supported by an
additional isolating integral of motion that forces the particles to
have a 1-D intersection with the orbital plane at a fixed binary
phase.

The equations of motion for the binary system are solved in terms of
the eccentric anomaly $\psi$ (Goldstein, Poole \& Safko 2002, Section
3.7). We use units where the gravitational constant $G$, the binary
semi-major axis $a$, and its total mass $m_1 + m_2$ are set to unity
so that, the binary period is $2 \pi$, and its frequency $\omega =
1$. The separation between the stars at time $t$, measured from
periastron where the azimuthal angle $\theta=0$, is given by the
radius $r$,

\begin{eqnarray}
r =  a (1-e \cos \psi)\, ,  \label{eq1}\\  \omega t = (\psi - e \sin
 \psi)  \, ,\label{eqkepler}\\  \cos \theta = a (\cos \psi - e)/r .
\end {eqnarray}

The binary eccentricity, defined as $e=\sqrt{1-b^2/a^2}$ where $a$,
and $b$ are the semimajor and semiminor axes, and the mass ratio $q =
m_2 / (m_1 + m_2)$ are the only free parameters. We use an Adams
integrator (from the NAG fortran library) to follow the motion of a
test particle moving in the orbital plane of the two stars.  Kepler's
equation (\ref{eqkepler}) is solved with a tolerance of $10^{-9}$.

The equations of motion of the test particle are solved in an inertial
reference frame using Cartesian coordinates, with their origin at the
centre of mass of the binary.  All test particle trajectories are
launched when the binary is at periastron, with the two components
lying on the $x$-axis. The computation is halted if the particle runs
away, moving further than 10 times the semimajor axis from the centre
of mass, or if it comes within a distance of either star that results
in a high number of force computations, in general due to close
approaches to the stars.

To find stable {\it invariant loops}, for which the phase space
coordinates of our test particle, traces a 1-dimensional curve on
successive passes through periastron, we launch particles from a
chosen position along the $x$-axis joining the two stars at
periastron, and examine the iterates in some two-dimensional subspace,
such as the $x-y$ plane.  We plot the positions of the test particle
at each complete binary period, and adjust the starting velocity $v_y$
until the iterates converge on a one-dimensional curve.  In practice,
we look at the scatter along the radial direction for those iterates
that lie within a sector that spans a small angle ($5^\circ$) about
the x axis when viewed from the centre of mass of the system.  We
adjust the launch velocity $v_y$ until the radial scatter of
periastron positions of the test particle in the $5^\circ$ sector
drops within a threshold value. A value of $10^{-4} a$ is used for
circumbinary loops and $10^{-6} a$ for circumstellar loops. These
values are consistent with the numerical errors in the orbit
integration. For the majority of orbits, 10 points within the sector
suffice and no more than 5 attempts are necessary to identify a given
loop. While we are in a region of stable invariant loops, the required
launch velocity $v_y$ is a continuous function of the starting point
$x$.

The numerical strategy we employ to solve the problem will allow us to
find only stable invariant loops. Particles launched close to an
unstable loop would diverge and the code would not be able to find
this kind of loops. However, it is the stable orbits we are interested
in. We do not calculate all the possible loops, but restrict our
attention to those that are symmetric about the line joining the two
stars when they are at periastron.  When the binary orbit is circular,
these are exactly the closed periodic orbits of a circumbinary disk.
Although our figures show a set of discrete curves, invariant loops
form continuous families in the same way as periodic orbits. We show
only a few of the possible invariant loops, for clarity.

\section{Geometrical Characteristics of Circumbinary Discs}\label{geom}

Discs in multiple stellar systems have attracted attention because of
their high abundance, and also due to the interesting effects of the
interaction on the disc morphology, better studied every day with the
improvement of observations. Of particular interest are the
circumbinary discs that, because of their low density, are very
difficult to observe. However, their importance arises from the
possibility that these envelope discs might be feeding putative
circumstellar discs, or harbor protoplanets, planets or any kind of
debris.

As in Section~3 of Paper~I, the inner radius of the circumbinary disk
is set by the criterion that the stable loops exist and do not
intersect each other. If the binary eccentricity is not small
($e>0.1$) then the loops become unstable before they begin to
intersect each other; we find no more one-dimensional curves, or the
test particles fall toward the stars or go out of the system.

To characterize the geometry of circumbinary discs, we use the
coefficients of the Fourier expansion of the innermost loop:

\begin{eqnarray}
A_k=\frac{1}{N} \sum_{i=1}^{N} s(\phi_i) cos(k\phi_i),  \,
\nonumber\\
B_k=\frac{1}{N} \sum_{i=1}^{N} s(\phi_i) sin(k\phi_i) \, \label{LopEll},
\end {eqnarray}

\noindent where ($s_i, \phi_i$) are the polar coordinates of $N$
evenly spaced (in $\phi$) points along the innermost stable loop,
measured from the binary center of mass. The modulus $\sqrt{A_k^2 +
B_k^2}$ is used to determine the mean distance to the barycentre
($k=0$) and the lopsidedness ($k=1$).

More than one hundred simulations were included to calculate the fits
we present in the next two subsections that provide the main
geometrical characteristics of the discs. These simulations were
performed in the eccentricity interval $e=[0.0,0.9]$, and mass ratio
$q=[0.1,0.5]$ (equivalent to sample the whole range $q=[0.1,0.9]$
because of the symmetry in the definition of $q$). It is worth to
mention that we have not included in the fits values for $q<0.1$ since
the behavior of the radius at those extreme values of $q$, changes
abruptly and requires many more calculations. We will produce fits for
extreme $q$ values (as is the case of planets) in a further paper. The
loops technique, however, allows us to reach these extreme cases and
we present in this paper an example of it (Section \ref{observations}).

At any time, the invariant loops form closed curves, which deform as
the binary follows its orbit, and return to their original shape when
the binary returns to the same phase. But we found in Paper~I that in
practice even the circumstellar loops do not deform strongly, and the
circumbinary loops even less.  Thus we measure the properties of the
cirumbinary disks, and show their shapes in the figures, at the time
of periastron passage.

\subsection{Lopsidedness of circumbinary discs}
\label{loopell}

\begin{figure}
\includegraphics[width=84mm]{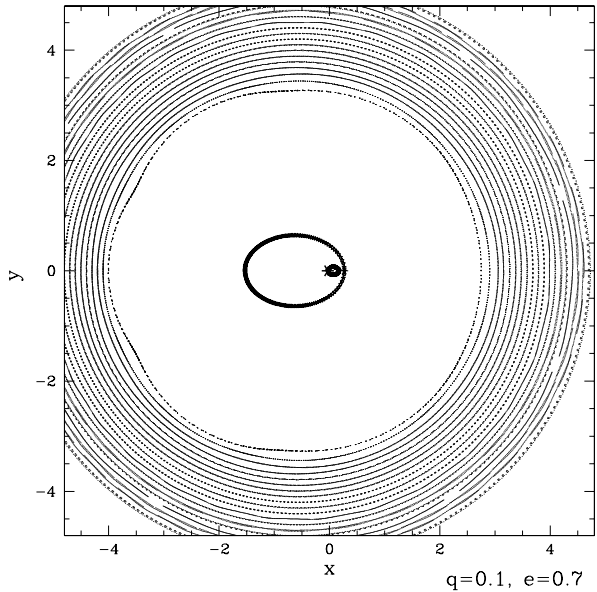}
\vspace{0.1cm}
\caption{Circumbinary disc computed with invariant loops for a binary
with $q=0.1$ and $e=0.7$, viewed at the moment of periastron marked
with the two stars at $(x,y)\approx (0,0)$.  The orbits of the stars
are shown with the darker curves. The disc is not centred about the
centre of mass of the system}
\label{fig.asymdisc}
\end{figure}

An interesting effect produced by a high binary eccentricity combined
with a large mass contrast, is the displacement of the geometric
centre of the circumbinary disc (inner edge) with respect to the
barycentre (See for example Figure \ref{fig.asymdisc}). This effect is
a physical characteristic that could explain some observed asymmetries
in discs (Duch\^ene \etal 2004; Boden \etal 2005; Kalas \etal 2005).
The displacement of the disc centre may affect calculations of the
disc's inclination (e.g. Itoh \etal 2002) since it is usual to link
asymmetries to inclination effects rather than to intrinsic
asymmetries in the geometrical centre of the discs.

In theoretical work of dynamics in planetary systems, there have been
several studies using linear analysis of perturbations, which consider
low mass ratios or low eccentricities (e.g. Wyatt \etal 1999, Kuchner
\& Holman 2003; Deller \& Maddison 2005). In these studies it is shown
how the presence of a second body (like a planet) in a system with an
eccentric orbit would impose a forced eccentricity on the orbits of
the constituent dust particles, thus shifting the geometric centre of
the disc away from the mass centre of the binary. In this manner, the
dust in slightly eccentric orbits would glow more brightly when it
approaches the pericentre. This could explain the asymmetry in the
double-lobed feature in many systems (Holland \etal 1998; Koerner
\etal 1998; Schneider \etal 1999; Telesco \etal 2000; Kalas \etal
2005; Freistetter 2007). In the same direction, Dermott \etal (1999)
find that if there is at least one massive perturber in the HR 4796
system that is on an eccentric orbit, then the system's secular
perturbations could cause the geometric centre of the disc to be
offset from the star.

\begin{figure}
\includegraphics[width=84mm]{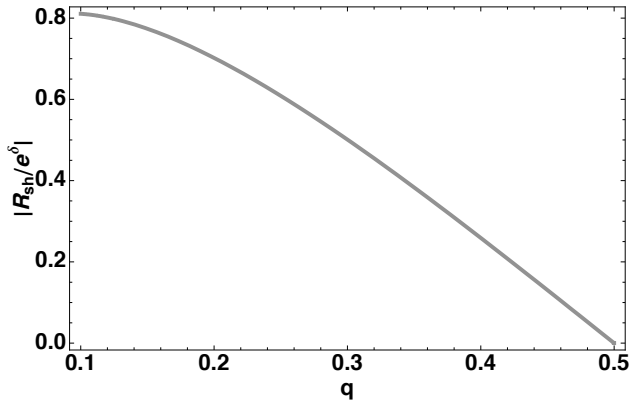}
\vspace{0.1cm}
\caption{Lopsidedness measured by the quantity $R_{sh}
(e,q)/e^\delta$, as a function of eccentricity $e$ and mass ratio $q$,
from equation \ref{eq.roff}}
\label{fig.Rshift}
\end{figure}

Our technique of invariant loops allows the displacement of the disk
to be calculated for arbitrary mass ratio and eccentricity. For mass
ratios $q \geq 0.1$ we have obtained a fit for the displacement of
the circumbinary geometric centre with respect to the centre of mass
of the binary (see Figure \ref{fig.Rshift}),

\begin{equation}
R_{sh} (e,q) =  -C_1 a e^\delta\ \left(0.5-q\right) \left[ q\left(1-q\right)\right]^\eta \, ,
\label{eq.roff}
\end {equation}

\noindent where $a$ is the semimajor axis $C_1=3.7$, $\delta=0.8$, and
$\eta=0.25$, give our best fit to the calculated off-centre
distance. The displacement in our calculations is always directed to
the left of the centre of mass due to the position we chose to place
the primary (left) and secondary (right) with respect to the centre of
mass of the binary at their pericentre (see Figure
\ref{fig.asymdisc}).

We find the best fit to the proposed equation by searching the minimum
residuals (some of them are given in Table \ref{table.diffshift})
produced by the square root of the sum of the squared differences
between the proposed function, equation \ref{eq.roff}, and the
computed displacements. The standard deviation comparing the
calculated data and the fit is 0.029$a$. The residual values in Table
\ref{table.diffshift} are in units of the semimajor axis, $a$.

\begin{table*}
\centering
\caption{Difference between the shift of the center of the
circumbinary disk computed using invariant loops, and the fit
(equation \ref{eq.roff}), in units of the semimajor axis, $a$ for some
chosen pairs ($e,q$).}
%\footnote{}\\
\begin{tabular}{@{}lrrrrrlrlr@{}}
\hline
$e$\ \ \ \ \ \ \ \ \  $q:$ & 0.1  & 0.2 & 0.3 & 0.4 & 0.5 \\
\hline
0.00  &  0.000 &  0.010 &  0.006 &  0.010 &  0.001 \\
0.20  & -0.005 & -0.025 &  0.030 &  0.009 & -0.004 \\
0.40  &  0.012 &  0.011 & -0.001 &  0.009 & -0.001 \\
0.60  & -0.012 & -0.052 &  0.012 &  0.024 &  0.000 \\
0.80  &  0.001 &  0.009 &  0.081 &  0.015 &  0.000 \\
\hline
\end{tabular}
\label{table.diffshift}
\end{table*}

The inner circumbinary rim is not exactly elliptical; close to
resonances, the loops can even become slightly triangular.  However,
unless the eccentricity is very close to zero, the shape of the inner
rim is close to circular and its maximum and minimum diameters lie
almost perpendicular. The eccentricity reaches a maximum value $e_d
\approx 0.15$, almost independent of the binary eccentricity.

\subsection{Inner Radii (the ``gap'')}\label{radii}

In Paper~I, we derived a simple relation for the size of circumstellar
discs as a function of binary mass ratio and eccentricity. In this
paper we extend the study to circumbinary discs and derive a relation
now for the inner radius (``gap''). Together, these radii define the
region where no loops exist.  We found that the change in radius of
the circumbinary disc with the binary phase is very small; so here we
calculate the inner radius of the circumbinary discs at one phase,
when the stars are at their periastron.

The best fit to our calculated inner radii is given by

\begin{equation}
R_{CB}(e,q) 
\approx C_2 a\left(1+\alpha e^\beta\right)\ \left(q(1-q)\right)^\gamma \,
 ,
\label{eq.rCB}
\end {equation}

\noindent where $a$ is the semimajor axis, $C_2=1.93$, $\alpha=1.01$,
$\beta=0.32$, and $\gamma=0.043$. The standard deviation obtained is
0.09$a$.

\begin{figure}
\includegraphics[width=84mm]{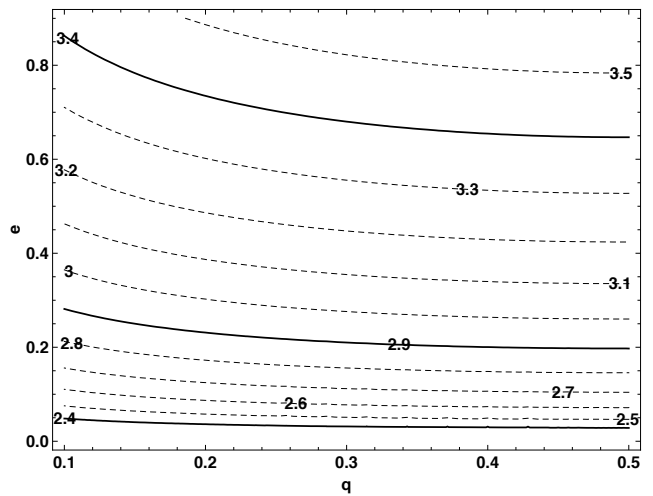}
\vspace{0.1cm}
\caption{Contour plot of the computed average inner radius $R_{CB}/a$
for the circumbinary discs, from equation \ref{eq.rCB}.}
\label{fig.Rcircumbinary}
\end{figure}

In Figure \ref{fig.Rcircumbinary}, we show a contour plot of the
approximation to the average radii for the circumbinary discs from
equation \ref{eq.rCB}. In Table \ref{table.res} we show the difference
between the relation \ref{eq.rCB}, and the computed radii.

\begin{table*}
\centering
\caption{Difference between the computed inner radii of the
circumbinary disks and the fit (from relation \ref{eq.rCB}), in units
of the semimajor axis, $a$, for some chosen pairs ($e,q$).}
%\footnote{}\\
\begin{tabular}{@{}lrrrrrlrlr@{}}
\hline
$e$ \ \ \ \ \ \ \ \ \ $q:$     & 0.1  & 0.2 & 0.3 & 0.4 & 0.5 \\
\hline

0.00  & -0.051 & -0.065  & -0.005  & -0.034  &  0.067  \\
0.20  &  0.000 & -0.060  & -0.016  &  0.107  & -0.204  \\
0.40  &  0.009 &  0.250  &  0.203  &  0.216  & -0.297  \\
0.60  &  0.170 & -0.022  &  0.106  &  0.105  & -0.202  \\
0.80  & -0.040 & -0.120  & -0.003  & -0.062  & -0.392  \\

\hline
\end{tabular}
\label{table.res}
\end{table*}

The ratio of equations \ref{eq.roff} and \ref{eq.rCB} gives an
estimate of the size of the displacement of the circumbinary disc
geometric centre, as a fraction of the inner circumbinary (average)
radius. In Table \ref{table.shiftperc} we present the ratio between
the calculated shifts and the corespondent average circumbinary
radius.

\begin{table*}
\centering
\caption{Displacements calculated from invariant loops for some cases
used for the fit in equation \ref{eq.roff}, given as a percentage of
the corresponding circumbinary average radius.}
%\footnote{}\\
\begin{tabular}{@{}lrrrrrlrlr@{}}
\hline
$e$\ \ \ \ \ \ \ \ \ $q:$ & 0.1 & 0.2 & 0.3 & 0.4 & 0.5 \\
\hline
0.00 &  0.000 &    0.557 &   0.320  &   0.538 &    0.051 \\
0.20 &  8.291 &    7.859 &   3.785  &   2.038 &    0.149 \\
0.40 & 12.340 &    9.682 &   7.177  &   3.363 &    0.035 \\
0.60 & 15.217 &   15.509 &   9.235  &   4.219 &    0.000 \\
0.80 & 20.179 &   17.178 &   9.577  &   5.776 &    0.000 \\
\hline
\end{tabular}
\label{table.shiftperc}
\end{table*}

\section{Comparison with Theoretical Work}
\label{theory}

In the problem of accretion and planet formation in binary systems, it
is of paramount importance to determine the regions where a
circumbinary disc can exist. Extensive literature exists, but mostly
devoted to the circular orbit binary case. Several studies have been
done about the dependence on mass ratio and orbital eccentricity for
the existence and characteristics of circumbinary discs. In
particular, Quintana \& Lissauer (2006) have simulated the late stages
of terrestrial planet formation within circumbinary discs in close
binary systems with a wide range of orbital eccentricities. Quintana
\& Lissauer (2007) simulated the final stages of terrestrial planet
formation in S- and P-type orbits within main-sequence binary star
systems.

In planetary studies a lot of work has been done (e.g. Holman \&
Wiegert 1999, hereafter HW99; Wyatt \etal 1999). In the last
reference, in particular, the authors address this issue by
investigating the long-term stability of planetary orbits numerically
using the eccentric restricted three-body problem. They launch
circular prograde orbits in the vicinity of the stars and in the
circumbinary region, looking for orbits that remain close to the stars
for more than 10,000 binary periods.  They provide a fit for the outer
radii of the circumstellar discs, and for the inner part of the
circumbinary disc, that depends only on the parameters of the binary
($a$, $e$, $q$). Their study has the advantage that it is able to
examine a full range of eccentricity and mass ratios. The
disadvantages are the lower precision of this technique, due to the
fact that a disc can live much more than the fiducial 10,000 binary
periods used by the authors to qualify an orbit as stable, and the
fact that it is expensive computationally.

In our work we identify stable non-intersecting loops where gas may
accumulate, a disc may develop and a planet may form.  This condition
plays the same role as searching for stable non-intersecting periodic
orbits in the potential of a circular binary; in either case, the
results could be modified by pressure or viscous forces.  In this
sense our search is more stringent than the study of HW99, who find
orbits where a planet may survive for long times around a binary, but
these orbits are permitted to intersect themselves or neighboring
orbits.  Still, a comparison of our results with those of HW99 is
relevant to gauge to what extent our different criteria result in
similar constraints.

\begin{figure}
\includegraphics[width=90mm]{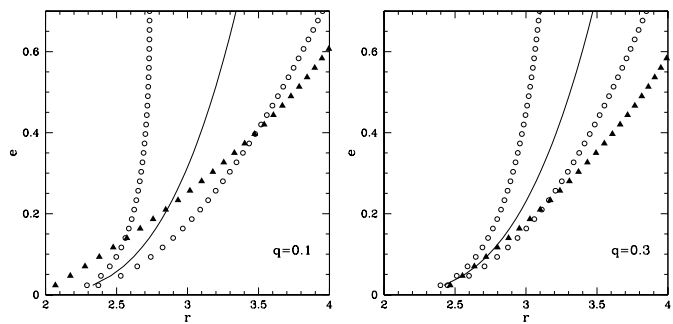}
\vspace{0.1cm}
\caption{Comparison between the radii calculated by the fit of HW99
(filled triangles), and the fit with invariant loops that gives the
mean radius (continuous lines) as eccentricity increases. We have also
included the maximum and minimum radius from the centre of mass due to
the shift of the disc (empty circles). Left panel: mass ratio $q=0.1$,
right: mass ratio $q=0.3$}
\label{fig.HW}
\end{figure}

We have calculated the difference between the fits made by HW99 for
the circumstellar and circumbinary discs. In the case of circumstellar
discs we find a good agreement with their results.  For the
circumbinary discs there are some differences.  In the Figure
\ref{fig.HW} we show the results of HW99 (filled triangles) and the
results obtained from invariant loops (continuous lines) of the
calculated average inner radii of circumbinary discs, including the
minimum and maximum distance from the center of mass (open circles),
vs the binary eccentricity, for two mass ratios, $q=0.1$ (left),
$q=0.3$ (right).  We see in the figure that the fit by HW99 gives in
general larger radii for the gap, especially at higher eccentricities
of the central binary.  For smaller eccentricities the radii provided
for the fit of HW99 are almost the same or even smaller than the ones
provided by the invariant loops.  It is likely that as the binary
becomes more eccentric, the phase space in which orbits can be trapped
so that they must remain close to the stable circumbinary loops
shrinks in volume.  That would make it less probable that the
initially-circular orbits of HW99 would lie in that trapped region.

\section{Application to Observations}\label{observations} 

Lim \& Takakuwa (2006) have already applied our results in Paper~I to
constrain the eccentricity of the binary in L1551 IRS5 to $e<0.3$,
based on the sizes of the circumstellar and circumbinary disks.

As a further application of our study, we have chosen four
systems. The first represents the prototype of circumbinary discs: GG
Tauri A. The second is UY Aurigae, the third is HD 98800 B, and the
last is Fomalhaut.

\subsection{The Circumbinary Disc of GG Tauri A System}\label{GGtau}
GG Tau is a well known young multiple system. The system has two
binary stars: GG Tau Aa/Ab, and GG Tau Ba/Bb. GG Tau A is an
interesting binary since it possesses a circumbinary disk resolved in
the millimeter wavelengths (Kawabe \etal 1993; Guilloteau \etal 1999).
It has been observed at high resolution in the optical (Krist \etal
2002), and in near-infrared (Roddier \etal 1996; Silber \etal 2000;
McCabe \etal 2002).

The structure of the circumbinary disc seems to be characterized by an
annulus with an inner radius between 180 and 190 $AU$ (Duch\^ene \etal
2004; Guilloteau \etal 1999), and an outer radius extending up to 800
$AU$. The total mass of the circumbinary material ($H_2$ + dust) is
$\approx0.12$ $M_\odot$ (Guilloteau \etal 1999).  Itoh \etal (2002)
derive an inclination of approximately $37^\circ$, assuming the orbit
of the binary is coplanar to the circumbinary disc.

The orbital characteristics of the central binary are still
controversial. While Roddier \etal (1996), propose an eccentric orbit
in which the stars are located near periastron, at the same
observation time, Krist \etal (2002) find they are close to apoastron
in a highly eccentric orbit. McCabe \etal (2002) deduce an elliptical
orbit $e=0.3\pm0.2$ with a semimajor axis of $a=35^{+22}_{-8}\
AU$. The mass of each component of GG Tau A a and b, binary is
obtained by White \etal (1999): $0.78\pm 0.1$ $M_\odot$, and $0.68\pm
0.03$ M$_\odot$ respectively (mass ratio $q\approx 0.47$).

\begin{figure}
\includegraphics[width=84mm]{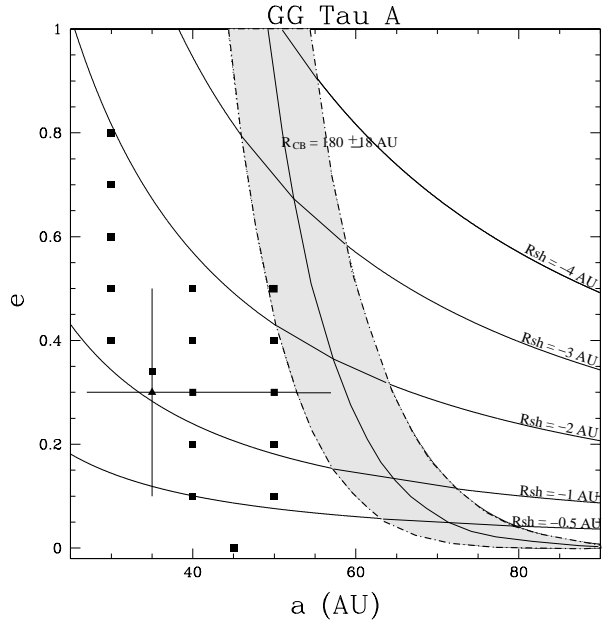}
\vspace{0.1cm}
\caption{Possible configurations for GG Tau A with our model using the
 observed inner radius of the circumbinary disc and an error from
 observations of $\pm 18$ $AU$ (shadowed region), assuming the disc is
 only sculpted by the binary. We have also included some observational
 and model predictions for the configuration. The filled squares are
 the predictions in the work of Itoh \etal 2002. The filled triangle
 is the prediction of McCabe \etal 2002 including the error
 bars. Finally we show five possible curves (continuous lines labeled
 with different $R_{sh}$) with the calculated shifts (equation
 \ref{eq.roff}), for the given $q=0.47$, and the corresponding pair
 $(a,e)$.}
\label{fig.evsaggtau}
\end{figure}

Based on equation \ref{eq.rCB}, we have calculated a band of possible
solutions for different semimajor axis and eccentricities that could
give an inner radius of $180\pm18$ $AU$ that we present in Figure
\ref{fig.evsaggtau}. In the same figure we locate the prediction by
McCabe \etal (2002), whose error bar in the semimajor axis locates it
inside our error zone (those with $e=0.3$ and semimajor axis around 55
$AU$). In the same manner, some of the values derived by the
predictions based in observations of Itoh \etal (2002), rest inside or
close to this region, specifically those with semimajor axis $a=50$
$AU$, and eccentricities $e=0.4,0.5$.

\begin{figure}
\includegraphics[width=84mm]{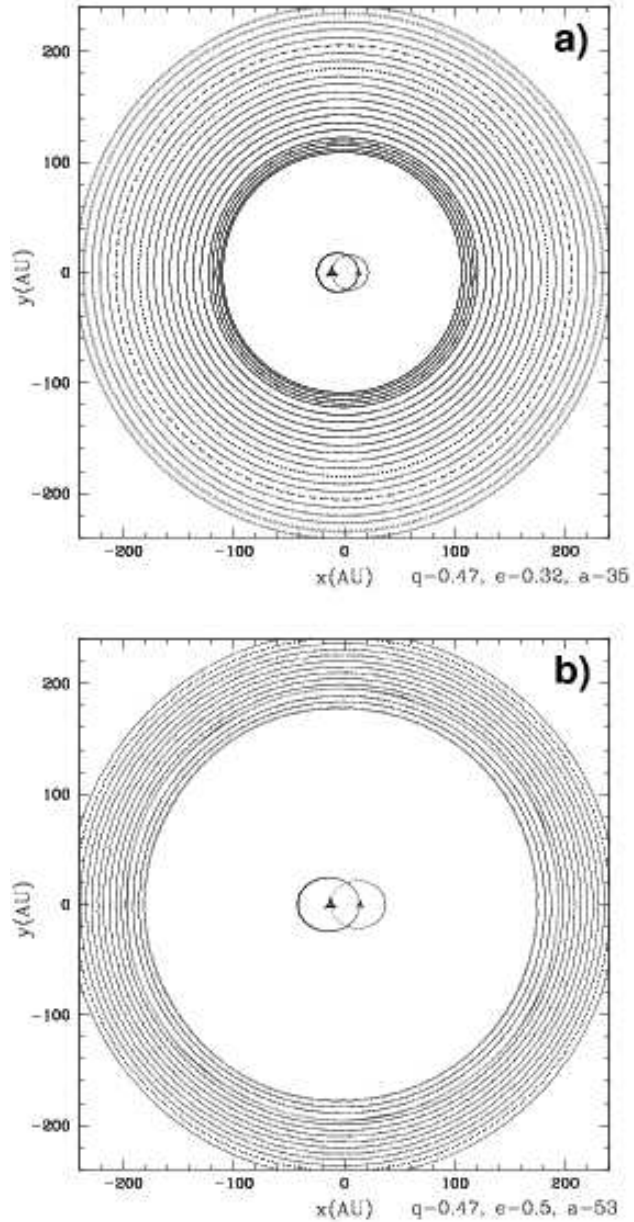}
\vspace{0.1cm}
\caption{a)Circumbinary disc computed with invariant loops for the
central value of eccentricity ($e=0.32$) and semimajor axis ($a=35$
$AU$) derived by McCabe \etal (2002) for a mass ratio $q=0.47$. b)
Same as a) but for a plausible invariant loops solution for
eccentricity (0.5) and semimajor axis ($a=53$ $AU$) that reproduces
the inner edge observed, of 180 $AU$, assuming that the only factor
sculpting the inner edge of the circumbinary disc is the main binary,
GG Tau A}
\label{fig.GGTAUloops}
\end{figure}

In figure \ref{fig.GGTAUloops} we present two possible configuration
for GG Tau constructed with invariant loops. The first plot is using
the values of eccentricity and semimajor axis reported by McCabe \etal
(2002) (triangle on Figure \ref{fig.evsaggtau}). Notice that this
configuration results in a significantly reduced inner gap which is
not compatible with the result reported by Duch\^ene \etal (2004) and
Guilloteau \etal (1999). In the second panel, we present a
configuration that is compatible with these two references
($R_{CB}\approx 180$~AU) and still within the error box of McCabe
around the preferred values of ($e=0.5$, $a=53$ $AU$). In both cases
we have assumed a mass ratio of $q=0.47$.  Although this parameter has
little importance in determining the circumbinary disc inner radius,
it is important in fixing its lopsidedness.

The circumbinary disc seems to be shifted away from the center of mass
of the binary (Duch\^ene \etal 2004) by $\sim 0''.16$ or $22$ $AU$ for
a distance of 140 pc.  We predict that the shift of the disc
geometrical centre with respect to the barycentre should be small
since the stars have almost the same mass.  As shown in Figure
\ref{fig.evsaggtau}, we find that the shift produced by the binary
system on the circumbinary disc, would be less than 4 $AU$.  We
calculate that a mass ratio of $q<0.3$ would be necessary to produce
both the observed shift and the observed inner radius of the
circumbinary disk. This is unlikely because the effective temperature
and luminosity of the two stars are very similar (White at al.1999).

Because the observational parameters determined by McCabe \etal (2002)
and Itoh \etal (2002) imply that the circumbinary disc should extend
much closer to the binary and the lopsidedness should be much smaller
than is predicted by considering orbits around the binary GG Tau A,
alternative theories have arisen. An interesting idea is the
possibility that the smaller GG Tau B binary (10\arcsec\ away and with
masses $0.12\pm 0.02$ and $0.044\pm 0.006$ $M_\odot$ for GG Tau B a
and b respectively) has something to do with the sculpting of the
circumbinary disc (Beust \& Deutrey 2006).  Unfortunately these
authors find that, although there are possible arrangements to explain
the outer edge of the disc, the binary B is not massive enough, and
not close enough to explain the wide inner radius in this way.

\subsection{The Binary System UY Aur}
\label{UYAur}

UY Aur is a binary system of classical T Tauri stars (Duch\^ene \etal
1999). It is located in the Taurus-Aurigae star forming region at an
approximate distance of 140 pc (Elias 1978). The projected separation
on the sky between UY Aur primary and secondary (A and B), is 120 $AU$
(Close \etal 1998). The spectral types of UY Aur A and B are estimated
to be M0 and M2.5, respectively (Hartigan \& Kenyon 2003).  Assuming a
circular orbit, the binary period is $\sim 1640\pm 90$ yr and the
total mass of the binary is $\sim 1.73\pm 0.29$ \msun\ (Hioki \etal
2007).

A circumbinary disc around the UY Aur binary was detected by
near-infrared, polarimetric and millimeter interferometric $^{13}$CO
emission observations (Duvert \etal 1998).  Its inner radius is about
520 $AU$ (Hioki \etal 2007).  These authors deproject the circumbinary
disk assuming that its inner edge is circular and the binary is
coplanar with it, to get a binary separation of 167 $AU$ and an
inclination of $42^o\pm3^o$. The disc seems to be not uniform showing
clumpy structure, circumstellar material inside the inner cavity and
an arm-like structure, probably created by accretion from the outer
region of the disc or stellar encounters (Hioki \etal 2007).

We calculated stable invariant loops in a system with the orbital
parameters proposed for the binary by Hioki \etal (2007), a binary
separation of 167 $AU$ and $e=0$.  We have set $q=0.5$, but Figure
\ref{fig.Rcircumbinary} shows that the edge of the CB disk does not
change significantly for any $q>0.1$.  This model would give an inner
radius for the circumbinary disc of no more than 340 $AU$, far less
than the observed 520 $AU$ for this system.

\begin{figure}
\includegraphics[width=84mm]{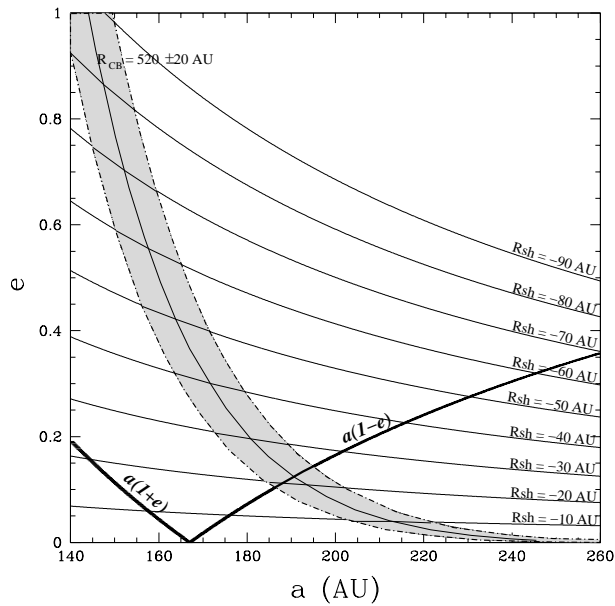}
\vspace{0.1cm}
\caption{Possible configurations for UY Aurigae with our fit for the
circumbinary radius for a mass ratio $q=0.25$, using the observed
circumbinary disc inner radius of 520 $AU$ and an error of $\pm 20$
$AU$ (shadowed region), assuming the disc is only sculpted by the
binary. We show nine possible curves (continuous lines labeled with
different $R_{sh}$) with the calculated shifts (equation
\ref{eq.roff}), for the given $q=0.25$, and the corresponding pair
$(a,e)$. The darker curves are the lines $a(1+e)=167$ $AU$ and
$a(1-e)=167$ $AU$}
\label{fig.uyaurq0.25}
\end{figure}

\begin{figure}
\includegraphics[width=84mm]{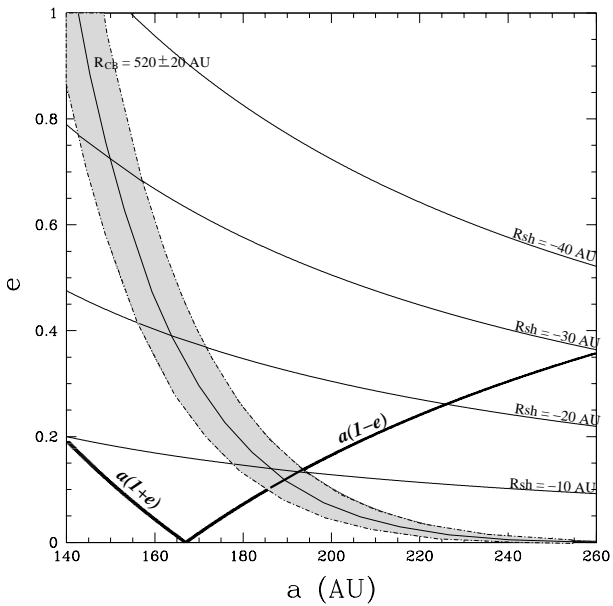}
\vspace{0.1cm}
\caption{Same as Figure \ref{fig.uyaurq0.25} but for $q=0.4$}
\label{fig.uyaurq0.4}
\end{figure}

We have proceeded by constructing a family of solutions as in the case
of GG Tau A (Section \ref{GGtau}).  This is, we provide a family of
solutions which reproduce the approximate inner radius of the
circumbinary disc. We construct two different possible configurations,
with different mass ratios, $q=[0.25,0.4]$ (Figures
\ref{fig.uyaurq0.25} and \ref{fig.uyaurq0.4}), which bracket $q=0.36$
as given by Hartigan \& Kenyon (2003). Although the inner radius of
the circumbinary disc is almost independent of the mass ratio (for $q
\geq 0.1$), the shift of the circumbinary disc increases as the mass
ratio becomes more unequal. The present separation is $167~AU$, which
must lie between the periastron and apastron separations. This
restricts $e, a$ to points above the curves given by $a(1-e)$ and
$a(1+e)$ in Figures \ref{fig.uyaurq0.25} and \ref{fig.uyaurq0.4}.
Thus the binary cannot be circular, but must have $e>0.1$ and for a
mass ratio of $q=0.36$ (Hartigan \& Kenyon 2003), the center of the
disk should be offset from the mass center of the stars in the
direction towards the center of the secondary star's orbit by
$R_{sh}\ga 0.05a$. Since the apparent separation has remained constant
since 1944 (Hioki \etal 2007), the system is unlikely to be very close
to periastron, implying a larger eccentricity.

\subsection{The Binary System HD 98800 B}
\label{HD98800B}

HD 98800 (HIP 55505, TWA 4A) is a hierarchical quadruple star system
in the TW Hya association. The separation between A and B components
is approximately $0.8''$ on a north-south line (Prato \etal
2001). Torres \etal (1995) find that both visual components are
themselves spectroscopic binaries. Tokovinin (1999) derived orbital
parameters $a=62~AU, e=0.5$ for the orbit of HD 98800 B around HD
98800 A.

The system HD 98800 B has excess flux in the mid-infrared, which was
interpreted by Soderblom \etal (1998) and by Prato \etal (2001) as a
circumbinary dust disk.  Their estimates of the inner radius of the
disk were 1.5 to 2 $AU$.  New observations of the stellar binary were
reported by Boden \etal (2005). They estimated visual and physical
orbits of the HD 98800 B subsystem with interferometric observations
combined with astrometric measurements by the Hubble Space Telescope
Fine Guidance Sensors. The orbital and physical parameters obtained in
that work are given in Table \ref{table.HD98800}.

\begin{table*}
\centering
\caption{Approximate Orbital Parameters of HD 98800 B from Boden \etal
(2005).}
%\footnote{}\\
\begin{tabular}{@{}ccrrrlrlr@{}}
\hline
$q$  & 0.45  \\
$e$  & 0.78    \\
$a$  & 0.98 $AU$ \\
Period (days)& 314 \\

\hline
\end{tabular}
\label{table.HD98800}
\end{table*}

Using the parameters inferred by Boden \etal (2005), Akeson \etal
(2007) used the results in our Paper~I to argue that the average inner
radius of the circumbinary disk should be at $R_{CB}\approx 3.4$ $AU$.
Thus the central hole in the cirumbinary disk should be significantly
larger than suggested by Soderblom \etal (1998) and Prato \etal
(2001).  Treating both binaries (A and B) as point masses, each with
the combined mass of its two stars, $q \approx 0.5$ and from Table 1
in Paper~I we predict that the maximum outer radius of the
circumbinary disc of B is about a tenth of the semi-major axis of the
orbit of A and B, or $6.2~AU$.  Akeson \etal compare these results
with a dynamical simulation of the system, in which the two stars of
binary B are followed separately while binary A is modeled as a single
star, and the circumbinary disk of B is represented by test particles.
After the equivalent of 1~Myr, the inner edge of the disk was at about
$3~AU$ and the outer edge at $10~AU$, in approximate agreement with
predictions from our invariant loops.

\begin{figure}
\includegraphics[width=84mm]{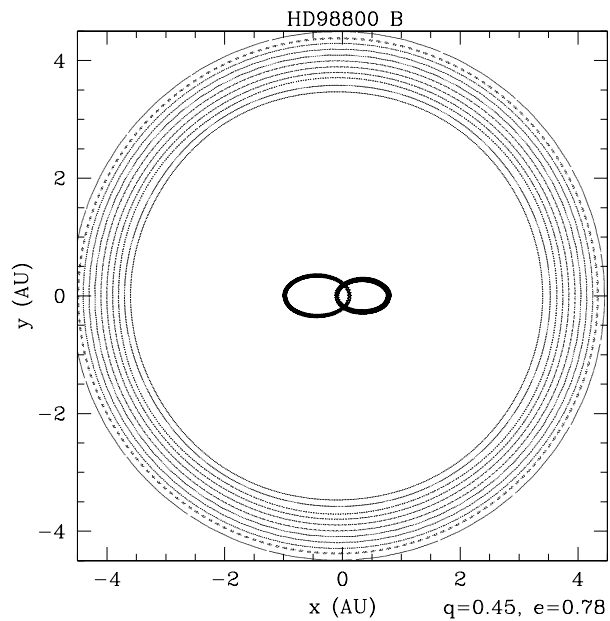}
\vspace{0.1cm}
\caption{Stable invariant loops making up the circumbinary disc of
HD98800 B, using the parameters of Boden \etal (2005), given in Table
\ref{table.HD98800}. The orbits of the stars are shown with the darker
curves}
\label{fig.HD98800}
\end{figure}

We have constructed the corresponding circumbinary disc for HD98800 B
(Figure \ref{fig.HD98800}).  Using equations \ref{eq.roff} and
\ref{eq.rCB}, we find that the disc center is shifted with respect to
the centre of mass of the system by $R_{sh}\approx -0.1$ $AU$. The
shift is small because the two stars have nearly equal mass.

\subsection{The Fomalhaut Dust Belt}
\label{Fomalhaut}

Fomalhaut (HD 216956 or $\alpha$ Pisces Austrinus) is a bright nearby
A3 V star of 2 $M_{\odot}$, an age of 200$\pm$100 Myr (Barrado \&
Navascues 1998) at a distance of 7.7 pc. It shows a dust (``debris'')
ring around it between 133 and 158 $AU$ from the central star (Aumann
1985; Guillet 1985; Dent \etal 2000; Holland \etal 1998, 2003) with an
approximate mass between 50 and 100 Earth mass (Kalas \etal
1995). This structure represents one of the best observed extrasolar
analog to our Kuiper belt. Fomalhaut's ring has an inclination of
$24^o$ away from edge on. The disc presents an asymmetry in the
brightness with the southern side nearer the star than the opposite
side (Stapelfeldt \etal 2004; Marsh \etal 2005). The deprojected
asymmetry (off-centred) of $\approx 15$ $AU$, the sharp inner cut at
133 $AU$, and the slight eccentricity of the disc $e_d\approx 0.1$ has
been studied recently by Kalas, Graham \& Clampin 2005 and Quillen
(2006) who have proposed that all the characteristics of this system
can be explained by the presence of a planet just interior to the ring
inner edge.

Several theories to form the sharp inner edge of the disc in Fomalhaut
have been proposed by different studies, the most accepted one until
this moment is the presence of inner planets.  The solution for
Fomalhaut is degenerate in the sense that several combinations of the
main parameters (mass ratios between the central star and the planet,
eccentricities and semimajor axes) could reproduce the observed values
for the inner radius of the dust disk $R_{CB}$, and the displacement
$R_{sh}$ of its center from the star's position. We explore here three
possible set of parameters given by other authors.

In Table \ref{fomalhaut_param} we show the selected parameters for
which we have calculated the invariant loops that would make up a
circumbinary disc, and their corresponding references.

\begin{table*}
\centering
\caption{Orbital parameters proposed for Fomalhaut. The first column
is the mass, the second is the planetary mass (in terms of the mass of
Jupiter, Saturn or Neptune), eccentricity $e$ and semimajor axis $a$
are in the third and fourth columns, the employed technique and
reference is given in the last two columns}
%\footnote{}\\
\begin{tabular}{@{}cccccccrlr@{}}
\hline
$M_{Fomal}$  & $M_{Pl}$  &   $e$   &   $a$  & Technique  & Reference\\
(M$_\odot$)     &         &         &   ($AU$) &  & \\
\hline
2.3  & 2 M$_J$ &   0.4   &  59 & N-body  & Deller \& Maddison (2005) \\
2    & 1 M$_S$ to 1 M$_N$ &   0.1   &  119 & Secular perturbations &  Quillen (2006) \\
\hline
\end{tabular}
\label{fomalhaut_param}
\end{table*}

\begin{figure}
\includegraphics[width=84mm]{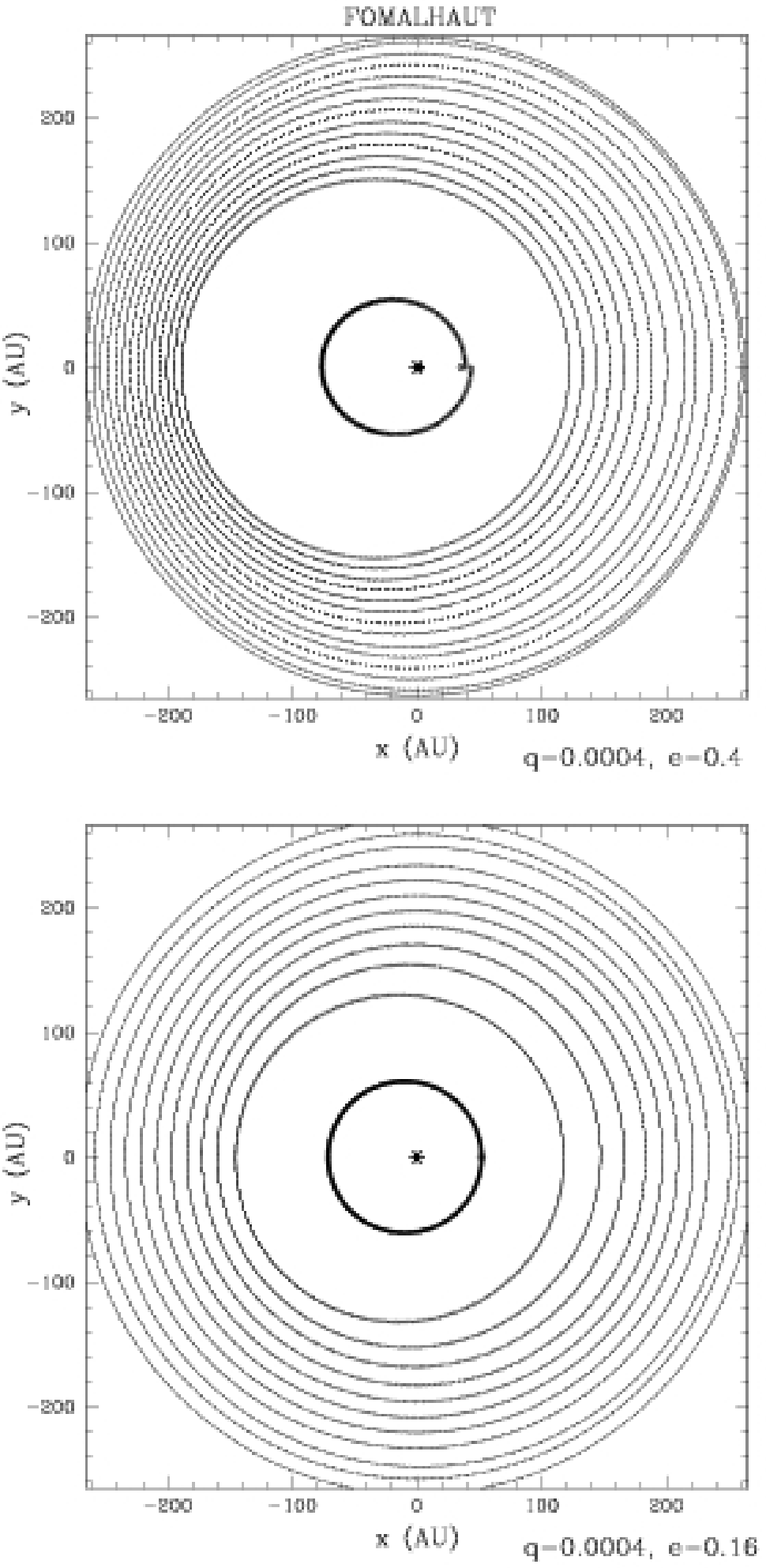}
\vspace{0.1cm}
\caption{Circumbinary disc of Fomalhaut. Upper panel: invariant loops
disc using the parameters ($e=0.4, a=59$~AU) in Table
\ref{fomalhaut_param} by Deller \& Maddison (2005). Lower panel: our
solution for the same planetary mass but with ($e=0.16, a=61.5$~AU).
The trajectory of the planet is shown with the darker curves.}
\label{fig.fomalhautDM}
\end{figure}

Under the assumption that a secondary body (a giant planet) is the
only source of asymmetry of the ring in Fomalhaut we have calculated
sets of invariant loops corresponding to both models in Table
\ref{fomalhaut_param}.  In the upper panel of Figure
\ref{fig.fomalhautDM} are given the results with invariant loops for
the approximation of Deller \& Maddison (2005). From Figure
\ref{fig.fomalhautDM} and Table \ref{fomalhaut_solution} we show that
the inferred value of eccentricity from Deller \& Maddison (2005) is
considerably larger than it should be to obtain the observed inner
radius of Fomalhaut's disk, and the shift, if the mass of the planet
is 2 M$_J$. In the lower panel of Figure \ref{fig.fomalhautDM} and
second row of Table \ref{fomalhaut_solution}, we show a solution for a
planet mass of 2 M$_J$ for which $R_{CB}$ and $R_{sh}$ are closer to
the observed values for Fomalhaut. We have also computed the
eccentricity $e_d$ of the inner rim of the circumbinary disk presented
in the last column of Table \ref{fomalhaut_solution}.

Quillen (2006) proposed solutions with a less-massive planet, using
the formulae of Wyatt \etal (1999) to calculate the expected
eccentricity.  For two extreme values of the range given in that work,
this is, the mass of Saturn and the mass of Neptune, given in Table
\ref{fomalhaut_param}, we have calculated the invariant loops to find
the average and the shift radii. The values given by Quillen (2006)
are close to the results that we derive from the invariant loops, as
can be seen in Table \ref{fomalhaut_solution}. This shows that in the
limit of small eccentricity, calculations based on invariant loops
agree with those from earlier methods which are valid only in those
regimes.

\begin{table*}
\centering
\caption{Disc characteristics from invariant loops for Fomalhaut
assuming the proposed masses of Deller \& Maddison (2005) and Quillen
(2006) for the planet, and our best approximation to the solution
taking the proposed masses by these authors. The first column shows
the assumed mass for Fomalhaut, the second is the proposed mass for
the planet around Fomalhaut (J=Jupiter, S=Saturn, N=Neptune), the
third and fourth are the eccentricity and the semimajor axis of the
planetary orbit, the fifth and sixth are the shift radii calculated
with invariant loops, the seventh column is the value of the
circumbinary inner rim eccentricity ($e_d$), and the last column
indicates the author of the paper we have taken the parameters from or
our best approximation to the parameters that approach the most to the
observed values of Fomalhaut.}
%\footnote{}\\
\begin{tabular}{@{}llllllllllllr@{}}
\hline
$M_{Fomal}$     & $M_{Pl}$  &   $e$   &   $a$   & $R_{sh}$ & $R_%F{
{CB}$ & $e_d$ & Work \\
(M$_\odot$)&           &         &   ($AU$)&   ($AU$)&   ($AU$) & &  \\
\hline
2.3& 2 M$_J$       & 0.4   & 59   & 36   & 155   &      & Deller \& Maddison (2005)\\
2.3& 2 M$_J$       & 0.16  & 61.5 & 16   & 133   & 0.15 & Our best approx. \\
2  & M$_N$ & 0.1   & 119  & 12.5 & 142.5 &      & Quillen(2006)    \\
2  & M$_N$ & 0.13  & 90.5 & 14.5 & 136   & 0.13 & Our best approx. \\
2  & M$_S$  & 0.1   & 119  & 12.5 & 148   &      & Quillen(2006)    \\
2  & M$_S$  & 0.13  & 106.4& 14   & 135   & 0.13 & Our best approx. \\
\hline
\end{tabular}
\label{fomalhaut_solution}
\end{table*}

\section{Conclusions}\label{conclusions}

We have extended the studies started in Paper~I to a more detailed
analysis of circumbinary discs in eccentric binary systems from the
geometrical and physical point of view.  The disks are defined by a
family of stable {\it invariant loops}, the analogs to stable periodic
orbits around a circular binary, which do not cross each other or
themselves.  Thus they define paths that can be followed by clouds of
gas, which dissipate energy when they run into each other.  Just as
with a binary in circular orbit, the circumbinary disk is truncated at
its inner edge when there are no longer any stable non-crossing orbits
for the gas to follow.  We have used this property of the invariant
loops to define the inner edge of the circumbinary disk.

We showed already in Paper~I that the inner radius of a circumbinary
disk depends strongly on eccentricity, opening wider gaps for higher
eccentricities.  The size of the inner hole depends only slightly on
the mass ratio.  The geometric centre of the circumbinary disc is
off-centre with respect to the centre of mass of the binary system.
The disc is closer to being symmetrical around the whole orbit of the
secondary star.

Here, we have explored the range of parameters to quantify both the
off-centering of the circumbinary disc with respect to the centre of
mass of the system, and the average inner radius of circumbinary
discs, as a function of mass ratio and eccentricity.  We compare our
results with the work of HW99 who searched for initially-circular
orbits that survive more than 10,000 periods of the binary.  When the
eccentricity is small this procedure gives similar results to ours,
but at larger eccentricity HW99 find fewer stable orbits close to the
binary, and hence larger inner gaps.  This could be related to the
fact that the larger the eccentricity, the smaller the available phase
space of orbits that are trapped so that they must remain close to a
stable invariant loop.

If the properties of a circumbinary disc are observable, it is
possible to constrain the binary system properties by comparing the
predictions based on invariant loops with what is observed. Likewise,
if the orbital parameters of a binary are known, the geometry of the
circumbinary and circumstellar regions permitted for stable orbits are
readily obtained.

For the well known circumbinary disc of the binary system GG Tau A, we
use the inner radius of the circumbinary disk to restrict the
possibilities for the binary parameters.  Since two stars have nearly
equal masses, we would expect the ring to be nearly symmetrical about
the center of mass of the stars.  The observed offset is substantial
and much larger than can be explained by orbital dynamics; effects
such as the finite ring thickness may be important (e.g. Duch\^ene
\etal 2004).

In the case of UY Aurigae, we show that the observed radius of the
circumbinary disk requires that the binary orbit be noncircular, with
eccentricity $ e > 0.1$.  The center of the disk should be offset from
the mass center of the stars in the direction towards the center of
the secondary star's orbit by $R_{sh}\ga 0.05a$.

We have modeled the system HD 98800 B with the parameters given by
Boden \etal (2005): the shift of the disc with respect to the centre
of mass of the system should be about $0.1~AU$.

Although the fits we provide here are valid for values of $q\ge 0.1$,
our technique is also applicable to extreme cases of $q \le 0.001$.
For the disk around Fomalhaut, we have used invariant loops to propose
plausible solutions for the orbital parameters of a planet that
explains the morphology of the debris disc of this system.

We reach similar results to Quillen (2006), but for the larger
planetary mass proposed by Deller \& Maddison (2005) we find that the
planet's orbit must be less eccentric.

\section*{Acknowledgments}
B. P. acknowledges Antonio Peimbert for useful discussions and
acknowledges projects CONACYT through grant 50720 and UNAM through
grant PAPIIT IN119708.

\label{lastpage}


\begin{thebibliography}{}

\bibitem[\protect\citeauthoryear{Akeson \etal}{2007}]{A07}
Akeson, R. L.; Rice, W. K. M.; Boden, A. F.; Sargent, A. I.;
Carpenter, J. M.; Bryden, G, 2007, ApJ 670, 1240

\bibitem[\protect\citeauthoryear{Barrado y Navascues}{1998}]{BN98}
Barrado y Navascues, D. 1998, A\&A, 339, 831

\bibitem[\protect\citeauthoryear{Bate}{1997}]{B97}
Bate, M. R., 1997, MNRAS, 285, 16

\bibitem[\protect\citeauthoryear{Bate \& Bonnell}{1997}]{BB97}
Bate, M. R. \& Bonnell, I. A. 1997, MNRAS, 285, 33

\bibitem[\protect\citeauthoryear{Beust \& Deutrey}{2006}]{BD06}
Beust, H.; Dutrey, A. 2006, A\&A, 446, 137

\bibitem[\protect\citeauthoryear{Boden \etal}{2005}]{BSA05}
Boden, Andrew F.; Sargent, Anneila I.; Akeson, Rachel L.;
Carpenter, John M.; Torres, Guillermo; Latham, David W.;
Soderblom, David R.; Nelan, Ed; Franz, Otto G.; Wasserman,
Lawrence H., 2005, ApJ, 635, 442

\bibitem[\protect\citeauthoryear{Bodenheimer \etal}{2000}]{BHL00}
Bodenheimer, Peter; Hubickyj, Olenka; Lissauer, Jack J. 2000, Icar,
143, 2

\bibitem[\protect\citeauthoryear{Bonnell \& Bastien}{1992}]{BB92}
Bonnell, I. \& Bastien, P. 1992, IAU Colloquium 135, 32, 206

\bibitem[\protect\citeauthoryear{Carpenter}{2000}]{C00}
Carpenter, John, M. 2000, AJ, 120, 3139

\bibitem[\protect\citeauthoryear{Close \etal}{1998}]{C98}
Close, L. M.; Dutrey, A.; Roddier, F.; Guilloteau, S.; Roddier, C.;
Northcott, M.; Menard, F.; Duvert, G.; Graves, J. E.; Potter, D.
1998, ApJ, 499, 883

\bibitem[\protect\citeauthoryear{Deller \& Maddison}{2005}]{DM05}
Deller, A. T.; Maddison, S. T. 2005, ApJ, 625, 398

\bibitem[\protect\citeauthoryear{Dermott \etal}{1999}]{DG99}
Dermott, S. F.; Grogan, K.; Holmes, E.; Kortenkamp, S. 1999, Formation
and Evolution of Solids in Space, Edited by J. Mayo Greenberg and
Aigen Li. Kluwer Academic Publishers, 1999., p.565

\bibitem[\protect\citeauthoryear{Duch\^ene \etal}{2004}]{DM91}
Duch\^ene, G.; McCabe, C.; Ghez, A. M.; Macintosh, B. A., 2004,
ApJ, 606, 969

\bibitem[\protect\citeauthoryear{Duch\^ene \etal}{1999}]{DM99}
Duch\^ene, G.; Monin, J.-L.; Bouvier, J.; Menard, F. 1999, A\&A, 351,
954

\bibitem[\protect\citeauthoryear{Duvert \etal}{1998}]{D98}
Duvert, G.; Dutrey, A.; Guilloteau, S.; Menard, F.; Schuster, K.;
Prato, L.; Simon, M. 1998, A\&A, 332, 867

\bibitem[\protect\citeauthoryear{Elias}{1978}]{E78}
Elias, J. H. 1978, ApJ, 224, 857

\bibitem[\protect\citeauthoryear{Els \etal}{2001}]{ESM01}
Els, S. G.; Sterzik, M. F.; Marchis, F.; Pantin, E.; Endl, M.;
K\"urster, M. 2001, A\&A, 370, L1

\bibitem[\protect\citeauthoryear{Fischer \etal}{2008}]{FM08}
Fischer, Debra A.; Marcy, Geoffrey W.; Butler, R. Paul; Vogt, Steven
S.; Laughlin, Greg; Henry, Gregory W.; Abouav, David; Peek, Kathryn
M. G.; Wright, Jason T.; Johnson, John A.; McCarthy, Chris; Isaacson,
Howard. 2008, ApJ, 675, 790

\bibitem[\protect\citeauthoryear{Freistetter \etal}{2007}]{FK07}
Freistetter, Florian; Krivov, Alexander V.; Lohne, Torsten. 2007,
A\&A, 466, 389

\bibitem[\protect\citeauthoryear{Goldstein \etal}{2002}]{G02}
Goldstein, H., Poole, C. \& Safko, J.  2002, {\it Classical
Mechanics}, (3d ed.; Addison Wesley)

\bibitem[\protect\citeauthoryear{Guilloteau \etal}{1999}]{GD99}
Guilloteau, S.; Dutrey, A.; Simon, M. 1999, A\&A, 348, 570

\bibitem[\protect\citeauthoryear{Haghighipour}{2006}]{H06}
Haghighipour, Nader, 2006, ApJ, 644, 543

\bibitem[\protect\citeauthoryear{Hartigan \etal}{2003}]{HK03}
Hartigan, Patrick; Kenyon, Scott J. 2003, ApJ, 583, 334

\bibitem[\protect\citeauthoryear{Hatzes \etal}{2003}]{H03}
Hatzes, Artie P.; Cochran, William D.; Endl, Michael; McArthur,
Barbara; Paulson, Diane B.; Walker, Gordon A. H.; Campbell, Bruce;
Yang, Stephenson. 2003, ApJ, 599, 1383

\bibitem[\protect\citeauthoryear{Hioki \etal}{2007}]{H07}
Hioki, Tomonori; Itoh, Yoichi; Oasa, Yumiko; Fukagawa, Misato; Kudo,
Tomoyuki; Mayama, Satoshi; Funayama, Hitoshi; Hayashi, Masahiko;
Hayashi, Saeko S.; Pyo, Tae-Soo; Ishii, Miki; Nishikawa, Takayuki;
Tamura, Motohide, 2007, AJ, 134, 880

\bibitem[\protect\citeauthoryear{Holland \etal}{1998}]{H98}
Holland, Wayne S.; Greaves, Jane S.; Zuckerman, B.; Webb, R. A.;
McCarthy, Chris; Coulson, Iain M.; Walther, D. M.; Dent, William
R. F.; Gear, Walter K.; Robson, Ian. 1998, Nature, 392, 788

\bibitem[\protect\citeauthoryear{Holman \& Wiegert}{1999}]{H99}
Holman, M. J., Wiegert P. A. 1999, AJ, 117, 621

\bibitem[\protect\citeauthoryear{Itoh \etal}{2002}]{I02}
Itoh, Yoichi; Tamura, Motohide; Hayashi, Saeko S.; Oasa, Yumiko;
Fukagawa, Misato; Kaifu, Norio; Suto, Hiroshi; Murakawa, Koji; Doi,
Yoshiyuki; Ebizuka, Noboru; Naoi, Takahiro; Takami, Hideki; Takato,
Naruhisa; Gaessler, Wolfgang; Kanzawa, Tomio; Hayano, Yutaka; Kamata,
Yukiko; Saint-Jacques, David; Iye, Masanori. 2002, PASJ, 54, 963

\bibitem[\protect\citeauthoryear{Jensen \etal}{1996}]{J96}
Jensen, Eric L. N.; Koerner, David W.; Mathieu, Robert D. 1996, AJ,
111, 2431

\bibitem[\protect\citeauthoryear{Kalas \etal}{2005}]{KP05}
Kalas, Paul; Graham, James R.; Clampin, Mark, 2005, Natur, 435, 1067

\bibitem[\protect\citeauthoryear{Koerner \etal}{1998}]{K98}
Koerner, D. W.; Ressler, M. E.; Werner, M. W.; Backman, D. E. 1998,
ApJ, 503, 83

\bibitem[\protect\citeauthoryear{Krist \etal}{2002}]{K02}
Krist, John E.; Stapelfeldt, Karl R.; Watson, Alan M. 2002, ApJ, 570,
785

\bibitem[\protect\citeauthoryear{Kuchner \etal}{2003}]{KH03}
Kuchner, Marc J.; Holman, Matthew J. 2003, ApJ, 588, 1110

\bibitem[\protect\citeauthoryear{Lada \& Lada}{2003}]{LL03}
Lada, Charles J.; Lada, Elizabeth A. 2003, ARA\&A, 41, 57

\bibitem[\protect\citeauthoryear{Lim \& Takakuwa}{2006}]{LT06}
Lim, J.; Takakuwa, S., 2006 ApJ 653, 425

\bibitem[\protect\citeauthoryear{Looney \etal}{2006}]{LTF06}
Looney, Leslie W.; Tobin, John J.; Fields, Brian D. 2006, ApJ, 652,
1755

\bibitem[\protect\citeauthoryear{Lyne \etal}{1988}]{L88}
Lyne, A. G.; Biggs, J. D.; Brinklow, A.; McKenna, J.; Ashworth,
M. 1988, Natur, 332, 45

\bibitem[\protect\citeauthoryear{Maciejewski \& Sparke}{1997}]{MS97}
Maciejewski, W., Sparke L. S. 1997, ApJL, 484, 117

\bibitem[\protect\citeauthoryear{Maciejewski \& Sparke}{2000}]{MS00}
Maciejewski, W., Sparke L. S. 2000, MNRAS, 313, 745

\bibitem[\protect\citeauthoryear{McCabe \etal}{2002}]{MD07}
McCabe, C.; Duch\^ene, G.; Ghez, A. M. 2002, ApJ, 575, 974

\bibitem[\protect\citeauthoryear{Mathieu}{1994}]{M94}
Mathieu, R. D. 1994, ARA\&A, 32, 465

\bibitem[\protect\citeauthoryear{Mathieu \etal}{1995}]{M95}
Mathieu, Robert D.; Adams, Fred C.; Fuller, Gary A.; Jensen, Eric
L. N.; Koerner, David W.; Sargent, Anneila I. 1995, AJ, 109, 2655

\bibitem[\protect\citeauthoryear{Mathieu}{2000}]{M00}
Mathieu, R. D.,  Ghez, A. M., Jensen, E. L. N. \& Simon, M. 2000,
in Protostar and Planets IV, ed. V. Mannings, A. P. Boss \&
S. S. Russell (Tucson: Univ. Arizona Press) 731

\bibitem[\protect\citeauthoryear{Mathieu \etal}{1997}]{M97}
Mathieu, Robert D.; Stassun, Keivan; Basri, Gibor; Jensen, Eric L. N.;
Johns-Krull, Christopher M.; Valenti, J. A.; Hartmann, L. W. 1997, AJ,
113, 1841

\bibitem[\protect\citeauthoryear{Pfahl \& Muterspaugh}{2006}]{PM06}
Pfahl, Eric; Muterspaugh, Matthew. 2006, ApJ, 652, 1694

\bibitem[\protect\citeauthoryear{Pichardo \etal}{2005}]{PSA05}
Pichardo, B.; Sparke, L. S.; Aguilar, L. A. 2005, MNRAS, 359, 521

\bibitem[\protect\citeauthoryear{Prato \etal}{2001}]{PR01}
Prato, L.; Ghez, A. M.; Pi\~na, R. K.; Telesco, C. M.; Fisher, R. S.;
Wizinowich, P.; Lai, O.; Acton, D. S.; Stomski, P. 2001, ApJ, 549, 590

\bibitem[\protect\citeauthoryear{Quillen}{2006}]{Q06}
Quillen, Alice C. 2006, MNRAS, 372, 14

\bibitem[\protect\citeauthoryear{Quintana \& Lissauer}{2006}]{QL06}
Quintana, Elisa V.; Lissauer, Jack J. 2006, Icar, 185, 1

\bibitem[\protect\citeauthoryear{Quintana \& Lissauer}{2007}]{QL07}
Quintana, Elisa V.; Lissauer, Jack J. 2007, arXiv, 0705.3444

\bibitem[\protect\citeauthoryear{Raghavan \etal}{2006}]{R06}
Raghavan, Deepak; Henry, Todd J.; Mason, Brian D.; Subasavage, John
P.; Jao, Wei-Chun; Beaulieu, Thom D.; Hambly, Nigel C. 2006, ApJ, 646,
523

\bibitem[\protect\citeauthoryear{Roddier \etal}{1996}]{R96}
Roddier, C.; Roddier, F.; Northcott, M. J.; Graves, J. E.; Jim,
K. 1996, ApJ, 463, 326

\bibitem[\protect\citeauthoryear{Rudak \& Paczynski}{1981}]{RP81}
Rudak, B., Paczynski, B. 1981, Acta Astron, 31, 13

\bibitem[\protect\citeauthoryear{Schneider \etal}{1999}]{S99}
Schneider, Glenn; Smith, Bradford A.; Becklin, E. E.; Koerner, David
W.; Meier, Roland; Hines, Dean C.; Lowrance, Patrick J.; Terrile,
Richard J.; Thompson, Rodger I.; Rieke, Marcia 1999, ApJ, 513, 127

\bibitem[\protect\citeauthoryear{Sigurdsson \& Phinney}{1993}]{SP93}
Sigurdsson, S.; Phinney, E. S. 1993, 1993, ApJ, 415, 631

\bibitem[\protect\citeauthoryear{Sigurdsson \etal}{2003}]{S03}
Sigurdsson, Steinn; Richer, Harvey B.; Hansen, Brad M.; Stairs, Ingrid
H.; Thorsett, Stephen E. 2003, Sci, 301, 193

\bibitem[\protect\citeauthoryear{Silber \etal}{2000}]{SJ00}
Silber, Joel; Gledhill, Tim; Duch=C3=AAne, Gaspard; M=C3=A9nard, Fran=C3=A7=
ois. 2000,
ApJ, 536, 89

\bibitem[\protect\citeauthoryear{Soderblom \etal}{1998}]{S98}
Soderblom, D. R.; King, J. R.; Siess, L.; Noll, K. S.; Gilmore, D. M.;
Henry, T. J.; Nelan, E.; Burrows, C.  J.; Brown, R. A.; Perryman,
M. A. C.; Benedict, G. F.; McArthur, B. J.; Franz, O. G.; Wasserman,
L. H.; Jones, B. F.; Latham, D. W.; Torres, G.; Stefanik, R. P.  1998,
ApJ, 498, 385

\bibitem[\protect\citeauthoryear{Telesco \etal}{2000}]{T00}
Telesco, C. M.; Fisher, R. S.; Pi\~na, R. K.; Knacke, R. F.; Dermott,
S. F.; Wyatt, M. C.; Grogan, K.; Holmes, E. K.; Ghez, A. M.; Prato,
L.; Hartmann, L. W.; Jayawardhana, R. 2000, ApJ, 530, 329

\bibitem[\protect\citeauthoryear{Tokovinin}{1999}]{T99}
Tokovinin, A. A., 1999, AstL 25, 669

\bibitem[\protect\citeauthoryear{}{1995}]{T95}
Torres, Guillermo; Stefanik, Robert P.; Latham, David W.; Mazeh,
T. 1995 ApJ, 452, 870

\bibitem[\protect\citeauthoryear{Trilling \etal}{2007}]{T07}
Trilling, D. E.; Stansberry, J. A.; Stapelfeldt, K. R.; Rieke, G. H.;
Su, K. Y. L.; Gray, R. O.; Corbally, C. J.; Bryden, G.; Chen, C. H.;
Boden, A.; Beichman, C. A. 2007, ApJ, 658, 1289

\bibitem[\protect\citeauthoryear{Weaver \etal}{2006}]{W06}
Weaver, H. A.; Stern, S. A.; Mutchler, M. J.; Steffl, A. J.; Buie,
M. W.; Merline, W. J.; Spencer, J. R.; Young, E. F.; Young,
L. A. 2006, Natur, 439, 943

\bibitem[\protect\citeauthoryear{White \etal}{1999}]{W99}
White, R. J., Ghez, A. M., Reid, I. N., Schultz, G., 1999, ApJ, 520,
811

\bibitem[\protect\citeauthoryear{Wyatt \etal}{1999}]{WD99}
Wyatt, M. C.; Dermott, S. F.; Telesco, C. M.; Fisher, R. S.; Grogan,
K.; Holmes, E. K.; Pi\~na, R. K., 1999, ApJ, 527, 918

%\bibitem[\protect\citeauthoryear{}{}]{}

\end{thebibliography}
\end{document}